# AFM and Raman studies of topological insulator materials subject to argon plasma etching

Isaac Childres[ab*], Jifa Tian[ab], Ireneusz Miotkowski[ab], Yong Chen[abc*]

[a]*Department of Physics, Purdue University, West Lafayette, IN, 47907, USA*
[b]*Birck Nanotechnology Center, Purdue University, West Lafayette, IN, 47907, USA*
[c]*School of Electrical and Computer Engineering, Purdue University, West Lafayette, IN, 47907, USA*

[*]Address correspondence to ichildre@purdue.edu, yongchen@purdue.edu

# AFM and Raman studies of topological insulator materials subject to argon plasma etching

Plasma etching is an important tool in nano-device fabrication. We report a study on argon plasma etching of topological insulator materials $Bi_2Se_3$, $Bi_2Te_3$, $Sb_2Te_3$ and $Bi_2Te_2Se$ using exfoliated flakes (with starting thicknesses of ~100 nm) derived from bulk crystals. We present data mainly from atomic force microscopy (AFM) and Raman spectroscopy. Through AFM measurements, plasma exposure is observed to decrease the thickness of our samples and increase surface roughness (with height fluctuations reaching as large as ~20 nm). We extract an etching rate for each type of material. Plasma exposure also causes a widening (especially $E_g^2$) of the characteristic Raman peaks, , with no significant change in peak position. The overall Raman intensity is observed to initially increase, then decrease sharply after the samples are etched below ~20 nm in thickness. Our findings are valuable for understanding the effects of argon plasma etching on topological insulator materials.

## 1. Introduction

Topological insulator (TI) materials have received much attention recently in the scientific community because of their distinct properties and potentials in nano-electronic applications.[1] Topological insulators are 3 dimensional materials that are insulating in the bulk, but conductive on the surface, creating a 2 dimensional system with Dirac Fermions. Some of the most common examples of TI materials are $Bi_2Se_3$, $Bi_2Te_3$, $Bi_2Te_2Se$[2] and $Sb_2Te_3$

Argon plasma etching is a tool commonly used in conjunction with lithographic techniques to fabricate topological insulator thin films or flakes such as $Bi_2Se_3$ and $Bi_2Te_3$ into devices. Our goal here is to investigate what changes plasma etching will induce in topological insulator materials. This study is inspired by our previous study of oxygen plasma etching of graphene.[3] The oxygen plasma etching results in marked changes in the graphene, including the introduction of new defect-related Raman peaks (*e.g.* D and D'), pit formation in atomic force microscopy (AFM) profiles and significantly modified electronic transport properties (e.g. reduced conductivity and mobility as well as emergence of weak localization).[3] We hope to investigate the effects of the similar defect-generation method on the Raman spectra and AFM profiles of TI materials.

## 2. Methodology

High quality bulk topological insulator crystals $Bi_2Se_3$, $Bi_2Te_3$, $Sb_2Te_3$ and $Bi_2Te_2Se$ have been synthesized using the Bridgeman method.[4] Samples studied in this work are fabricated using micromechanical exfoliation of bulk crystals onto a 300 nm $SiO_2$/Si substrate. Once peeled, the samples are located using an optical microscope and then confirmed to be ~100 nm thick using AFM. Flakes with such thickness are desired so that they can be gradually etched away in a reasonable amount of argon plasma exposures in our study. Once topological insulator flakes are identified to be the proper thickness, they are then characterized further using AFM and Raman spectroscopy before and after various amounts of exposure to an argon plasma.

Our plasma etching is performed in a microwave plasma system (Plasma-Preen II-382) operating at 100 W. A constant flow of argon (3 sccm) is pumped through the sample space in rough vacuum (<40 Torr), and the gas is excited by microwaves (operating for 1 or 2 s at a time), generating an ionized argon plasma. The microwave-excited plasma pulses are applied to the samples cumulatively, and AFM and Raman measurements are performed in ambient atmosphere and temperature after each successive exposure.

AFM measurements are performed using an ambient AFM system (NT-MDT NTEGRA Probe NanoLaboratory) operating in the tapping mode, where an area of ~10 μm-by-10 μm around a TI flake is scanned. The thickness quoted below is averaged from an interior area of the sample.

Raman measurements are performed using a confocal Raman microscope (Horiba XploRA) with a 532 nm excitation laser, averaging 4 acquisitions of 20 seconds each. The power of the laser is kept low (~200 μW) with a laser spot size of ~1 μm in diameter. We use Raman spectroscopy to study the $E_g^2$ and $A_{1g}^2$ crystal vibrational modes[5] in our topological insulator samples. We cannot observe the lower vibrational modes $E_g^1$ and $A_{1g}^1$ because they appear at wavenumbers inaccessible for the filter of our Raman detector.

## 3. Results and discussion

Figures 1(a-c) show selected AFM images of a $Bi_2Se_3$ flake after various amounts of plasma etching. Height profiles along the (colored online) horizontal lines for various stages of the etching process are plotted in figure 1(d). As is the case of all the topological insulator flakes we studied, plasma etching reduces the height of the samples in a non-uniform manner across the whole flake. Initially, the flakes exhibit a relatively smooth surface. However, the surfaces become increasingly rough the longer they are plasma etched, reaching a roughness (height fluctuation) as large as ~20 nm. Additionally, some edges of a sample are etched at a much lower rate than the rest of the sample, resulting in large disparities in height across the sample after a long plasma etching time. Given enough time (~30 s for this sample), however, even the slowly etched edges of the samples are completely removed from the substrate.

Similar AFM measurements are performed on $Bi_2Te_3$ (Figure 2), $Bi_2Te_2Se$ (Figure 3) and $Sb_2Te_3$ after various amounts of plasma etching, with qualitatively similar findings.

Figure 4(a) shows the average interior thicknesses of 4 different types of topological insulators – $Bi_2Se_3$, $Bi_2Te_3$, $Sb_2Te_3$ and $Bi_2Te_3Se$ – as functions of plasma etching time. Here the height is averaged over an interior area of the sample to avoid less-etched edges. In our study, we find that the $Bi_2Se_3$, $Bi_2Te_3$ and $Bi_2Te_2Se$ samples are all etched at similar rates (~10-13 nm/s), while $Sb_2Te_3$ has a faster etching rate (~40 nm/s). We also calculate and plot the RMS of the interior heights, figure 4(b), which also indicate an increased surface roughness as etching time increases (except when the sample is almost etched away with height approaching that of the surrounding substrate to cause the RMS to decrease again).

Figure 5(a) shows the Raman spectra of our $Bi_2Se_3$ sample before and after exposure to a progression of argon plasma etchings. The initial Raman spectrum shows peaks at 130 cm$^{-1}$ ($E_g^2$) and 173 cm$^{-1}$ ($A_{1g}^2$), both of which agree well with the values reported in previous works.[5,6,7] We can also see the tail of the $A_{1g}^1$ peak – expected to appear at 72 cm$^{-1}$ – but the full peak is largely truncated by the filter. The positions, intensities and widths of the $E_g^2$ and $A_{1g}^2$ peaks, determined using Lorentzian fits to the spectra, are plotted in 5(b,c). We observe the positions of the $E_g^2$ and $A_{1g}^2$ peaks remain relatively constant throughout the etching process, seen in figure 5(b). We do, however, observe a moderate increase in the width of the $E_g^2$ peak with increased etching time, as also seen in figure 5(b). Figure 5(c) shows an increase in the intensity of both the $E_g^2$ and $A_{1g}^2$ peaks with increased etching before the intensity of the entire spectra decreases sharply as most of the flake is etched away. The intensity ratio $I(E_g^2)/I(A_{1g}^2)$ largely decreases as etching increases, as seen in figure 5(c).

We see similar trends in the progression of the Raman spectra of $Bi_2Te_3$, figure 6, and $Bi_2Te_2Se$, figure 7, as those trends observed in $Bi_2Se_3$. The initial Raman spectrum of $Bi_2Te_3$ shows peaks at 100 cm$^{-1}$ ($E_g^2$) and 132 cm$^{-1}$ ($A_{1g}^2$), which agree well with values from previous reports,[5,8,9] and the initial Raman spectrum of $Bi_2Te_2Se$ shows peaks at 101 cm$^{-1}$ ($E_g^2$) and 138 cm$^{-1}$ ($A_{1g}^2$). In the case of $Bi_2Te_3$, we note both the $E_g^2$ and $A_{1g}^2$ peak show a pronounced widening even for relatively small levels of plasma exposure.

We observe an increase in the Raman spectrum intensities of all our studied materials as plasma exposure increases, reaching the highest intensities when our sample thicknesses are reduced to ~20 nm (before the flake is etched away). Previous reports have also measured an increase in Raman intensity for sufficiently thin films. This intensity increase has been attributed to constructive interference of internal reflections of the Raman signal in the films.[6,10,11,12] We have also seen similar trends in Raman intensities as dependent on the thickness using un-etched TI flakes (see supplemental materials) consistent with those in previous reports. However, for our plasma etched samples, we see increases in intensity at thicknesses larger than those seen in previous reports.[6] We attribute this to increased roughness on the material surface causing areas of the sample reaching thickness lower than the average thickness.

We also note a decreasing trend in the intensity ratio $I(E_g^2)/I(A_{1g}^2)$ (particularly pronounced for $Bi_2Te_3$) as plasma etching increases. In a previous report, a reduction in this intensity ratio for thinner $Bi_2Te_3$ films is attributed to the emergence of infrared modes ($A_{1u}$ and $A_{2u}$, the latter of which has the same frequency as $A_{1g}^2$).[13] Infrared modes becoming Raman active is caused by a breaking of inversion symmetry, which has been attributed in thin TI films to be caused by a breaking of crystal symmetry or built-in electric fields on the surface of the material.[6,13] We note, however, in our case we do not see an emergence of the $A_{1u}$ (116 cm$^{-1}$) IR-activated peak in our plasma-etched samples. This could be related to doping in our samples causing an insufficient E-field on the surface or the defects in our samples.

The other significant trend in Raman spectra we observe is the increasing width of the $A_{1g}^2$ and $E_g^2$ peaks as plasma etching increases. Widening of the Raman peaks can be attributed to a decrease in phonon lifetime, caused by increased phonon scattering in the sample.[14] A previous report has noted an increase in Raman peak width in sufficiently thin TI films (<15 nm).[7] In this study, we see an increased width even for relatively thick films (~80 nm). We suggest that the disorder and surface roughness generated by plasma etching may play an important role in phonon scattering and Raman peak broadening in our samples.

To reinforce the evidence that this increased peak width is largely caused by plasma-induced disorder (rather than just a thickness reduction), we also measured the Raman spectra of pristine peeled $Bi_2Se_3$ and $Bi_2Te_3$ films for a wide range of thicknesses (figure S1) and saw no significant increase in $A_{1g}^2$ and $E_g^2$ peak width down to ~10 nm thickness. As an additional indication of the effect of disorder generated by plasma etching, we also measured the magneto-resistance (figure S2) of a thick (~100 nm) $Bi_2Se_3$ device before and after a brief amount of oxygen plasma etching that does not significantly reduce the thickness. After the etching, we see emergence of a small weak anti-localization feature, indicating increased electron scattering from defects generated by plasma etching.[15] A more systematic study of the electronic transport properties of plasma-etched TI samples is however beyond the scope of the current work, which focuses on AFM and Raman studies.

**4. Conclusions**

In summary, we have observed various significant effects due to plasma etching of TI materials. From AFM measurements, we see a decrease in sample thickness and an increase in surface roughness for increased plasma exposure. In our Raman measurements, we observe an increase in overall intensity, a decrease in the intensity ratio $I(E_g^2)/I(A_{1g}^2)$ and an increase in peak widths. We attribute these changes in Raman spectra both to decreased sample thickness and increased phonon scattering caused by argon plasma-generated defects and roughness.

**Supplemental Information**

Figure S1 shows Raman spectra on non-plasma-irradiated $Bi_2Se_3$ and $Bi_2Te_3$ samples of a variety of thicknesses (8-110 nm as measured by AFM) exfoliated from the same crystals as the irradiated samples. Below 20 nm, the samples show a marked increase in intensity, fitted to a reflectance model for thin films accounting for multireflection within the film (using a modified index of refraction for the films).[10] However, we observe no discernible upward trend with decreasing thickness (down to ~8 nm) for the FWHM of the $E_g^2$ and $A_{1g}^2$ peaks. This is different from Ref. 7, which documented an increase in the FWHM of the $E_g^2$ peak for $Bi_2Se_3$ nanoplatelets (fabricated using polyol synthesis) below 20 nm in thickness.

To see an extra indication of disorder in topological insulators due to argon plasma etching, we fabricate a device to measure the 4-terminal resistance of a ~100 nm thick exfoliated $Bi_2Se_3$ flake using gold electrodes created by electron-beam lithography. Figure S2 shows this resistance measured under a varying magnetic field before and after a small amount of argon plasma etching that did not significantly reduce the device's thickness. After argon plasma etching we see a small feature of reduced resistance centered at B=0 which we attribute to weak anti-localization.[13] This weak anti-localization feature indicates increased electron scattering due to irradiation-induced defects in the sample.

**Acknowledgements**

This work has been partially supported by the Mesodynamic Architectures program of the Defense Advanced Research Projects Agency. Yong P. Chen also acknowledges support from the Miller Family Endowment.

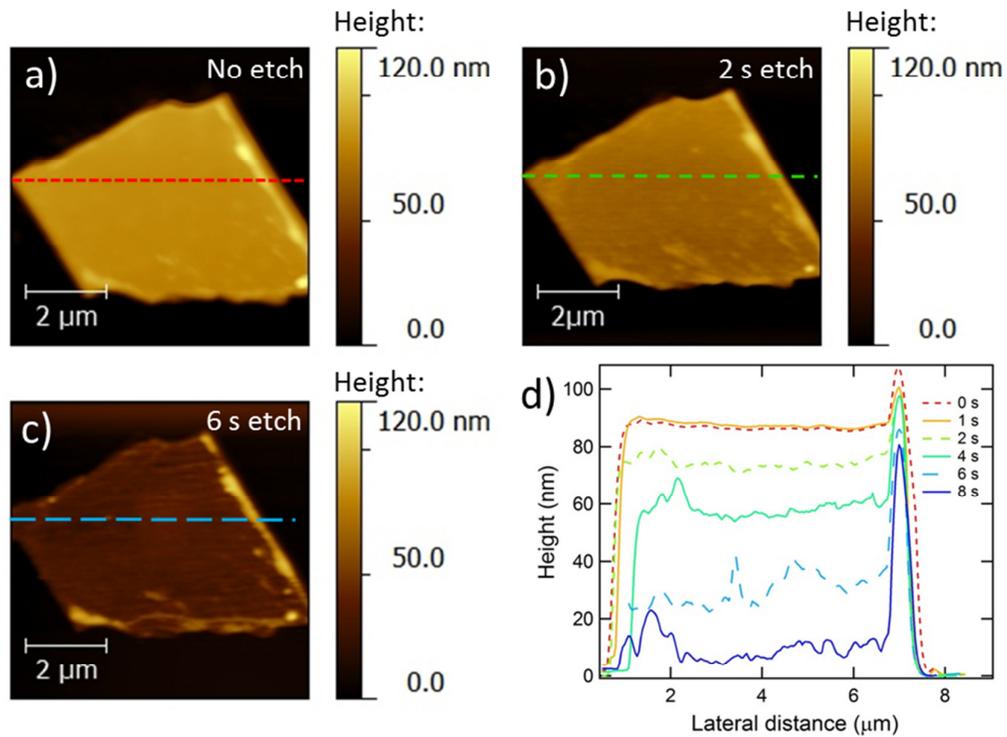

Figure 1. (a-c) Atomic force microscopy images taken of a $Bi_2Se_2$ flake (~86 nm thick) before (a) and after 2 s (b) and 6 s (c) of argon plasma etching. The thickness of the flake decreases with continued etching. (d) Measured height profiles along the dashed lines (color online) shown (a-c) for various etching times

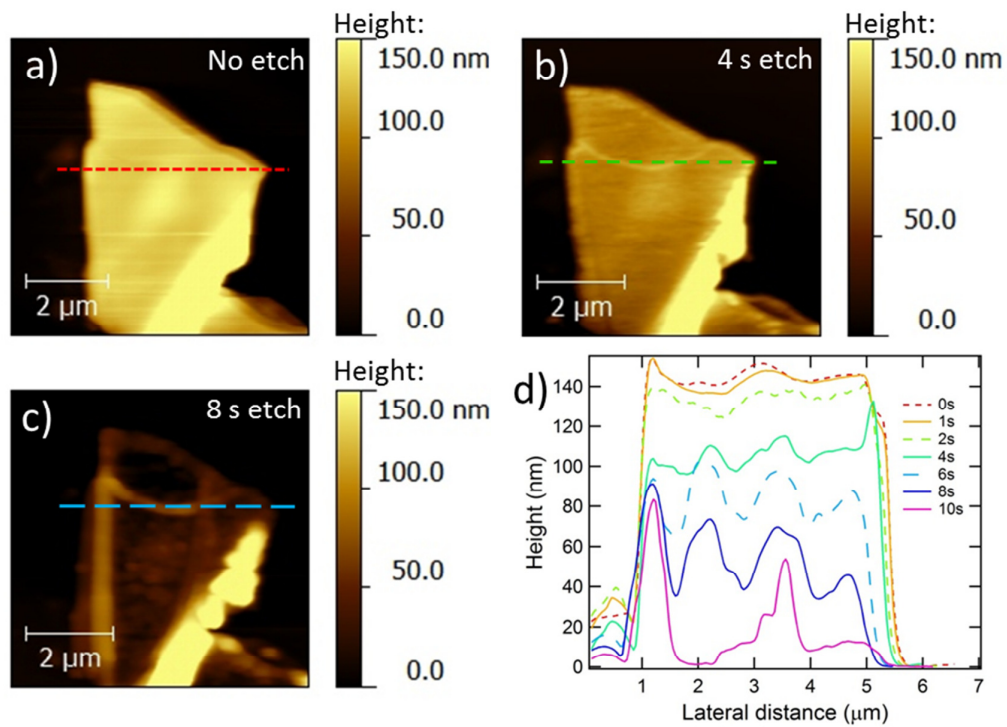

Figure 2. (a-c) Atomic force microscopy images taken of a $Bi_2Te_2$ flake (~130 nm thick) before (a) and after 4 s (b) and 8 s (c) of argon plasma etching. The thickness of the flake decreases with continued etching. (d) Measured height profiles along the dashed lines (color online) shown (a-c) for various etching times.

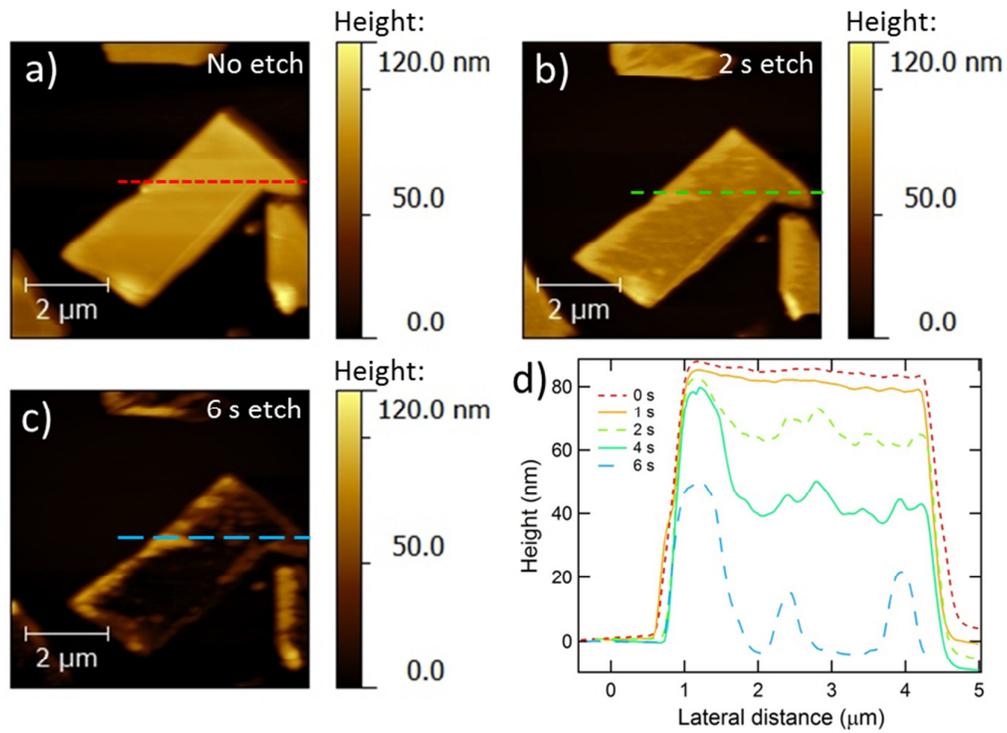

Figure 3. (a-c) Atomic force microscopy images taken of a $Bi_2Te_2Se$ flake (~94 nm thick) before (a) and after 2 s (b) and 6 s (c) of argon plasma etching. The thickness of the flake decreases with continued etching. (d) Measured height profiles along the dashed lines (color online) shown (a-c) for various etching times.

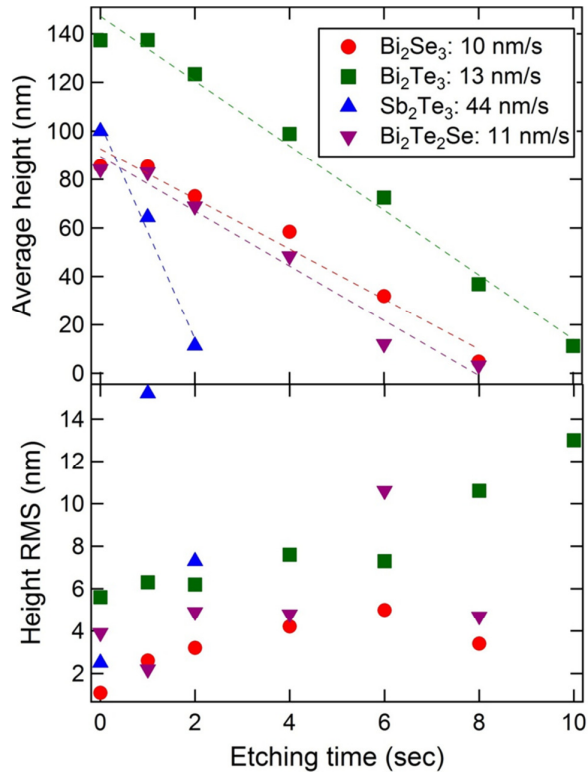

Figure 4. (a) Average sample height versus plasma etching time for 4 different types of topological insulator samples. Prior to etching, all samples were around 100 nm thick. The height is averaged from an interior area of the sample. The $Bi_2Se_3$, $Bi_2Te_3$ and $Bi_2Te_2Se$ samples are all etched at similar rates (~10-13 nm/s), while $Sb_2Te_3$ has a faster etching rate of ~40 nm/s in this test. (b) The RMS of the interior heights versus plasma etching time. The RMS increases as the etching increases, indicating increased surface roughness.

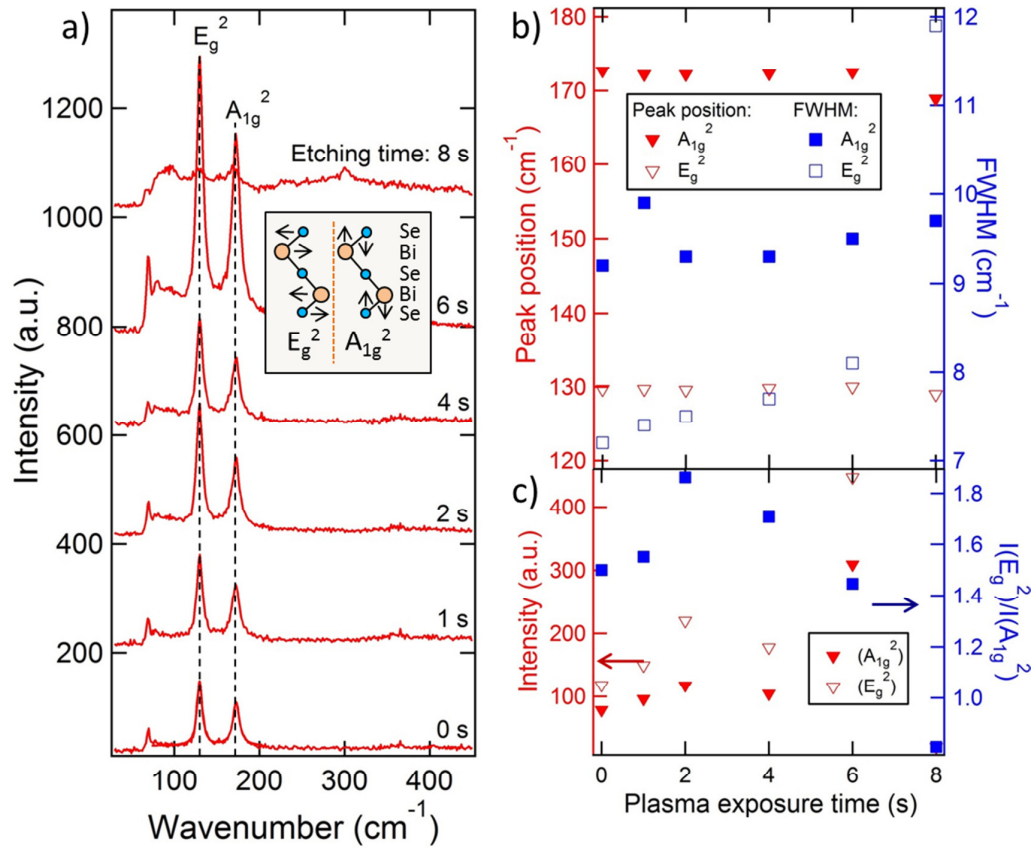

Figure 5. (a) Progression of Raman spectrum of a 86 nm-thick $Bi_2Se_3$ sample subjected to various amounts of argon plasma etching. The etching time is accumulated over a series of exposures. Spectra are offset vertically for clarity and are measured using a 532 nm excitation laser. The initial Raman spectrum shows peaks at ~130 cm$^{-1}$ and ~173 cm$^{-1}$, corresponding to the $E_g^2$ and $A_{1g}^2$ vibrational modes (inset). (b) The positions and full widths at half max of the $E_g^2$ and $A_{1g}^2$ peaks as a function of argon plasma exposure time. (c) The intensity of the $E_g^2$ and $A_{1g}^2$ peaks, as well as the intensity ratio $I(E_g^2)/I(A_{1g}^2)$ as a function of argon plasma etching time.

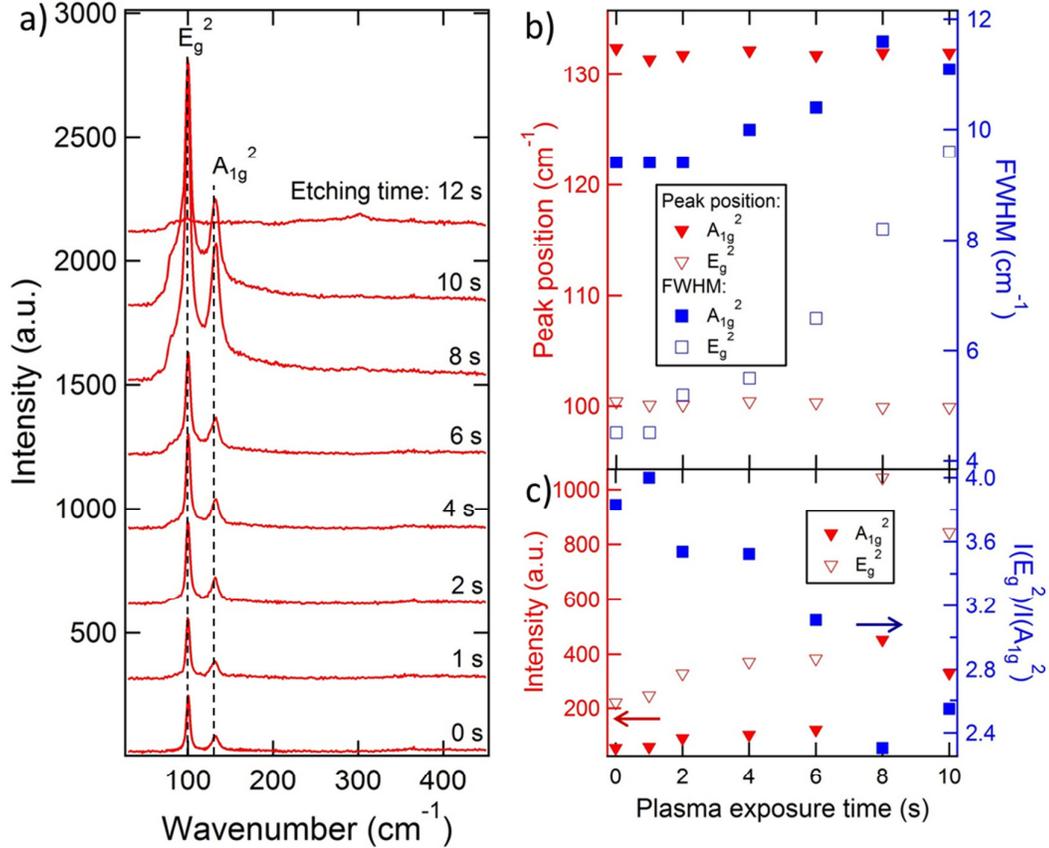

Figure 6. (a) Progression of Raman spectrum of a 130 nm-thick $Bi_2Te_3$ sample subjected to various amounts of argon plasma etching. The etching time is accumulated over a series of exposures. Spectra are offset vertically for clarity and are measured using a 532 nm excitation laser. The initial Raman spectrum shows peaks at ~100 $cm^{-1}$ and ~132 $cm^{-1}$, corresponding to the $E_g^2$ and $A_{1g}^2$ vibrational modes. (b) The positions and full widths at half max of the $E_g^2$ and $A_{1g}^2$ peaks as a function of argon plasma exposure time. (c) The intensity of the $E_g^2$ and $A_{1g}^2$ peaks, as well as the intensity ratio $I(E_g^2)/I(A_{1g}^2)$ as a function of argon plasma etching time.

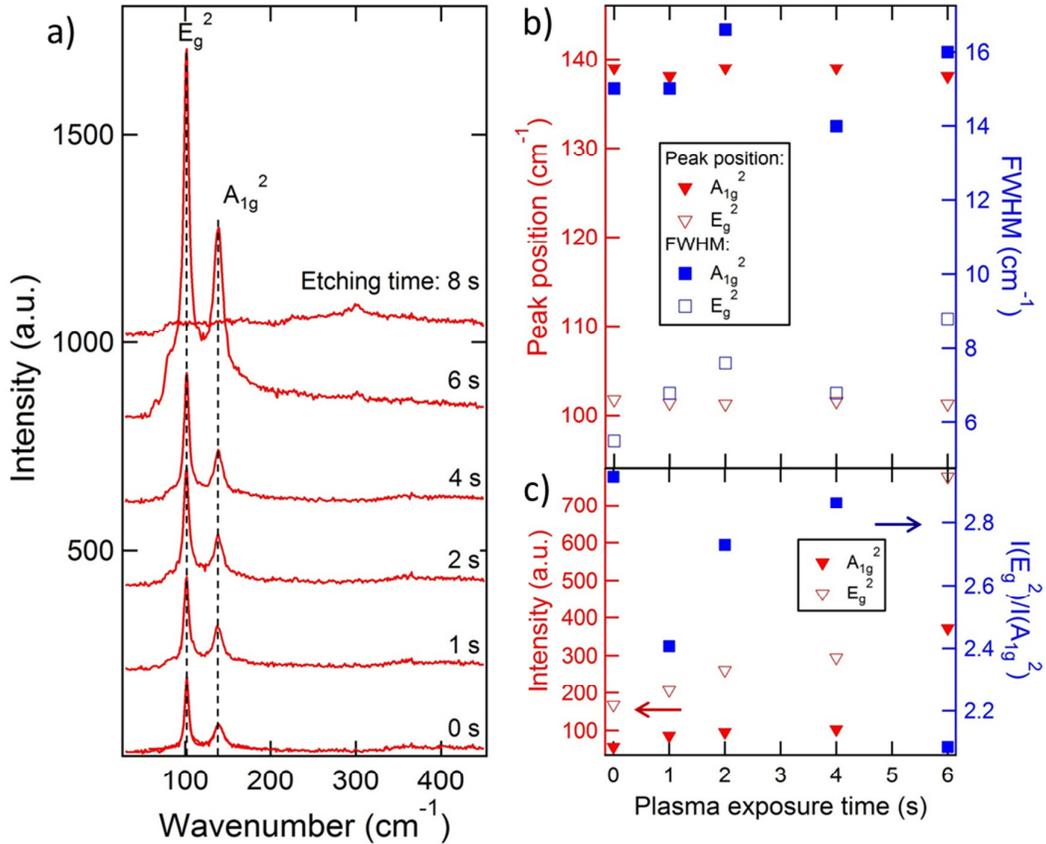

Figure 7. (a) Progression of Raman spectrum of a 94 nm-thick $Bi_2Te_2Se$ sample subjected to various amounts of argon plasma etching. The etching time is accumulated over a series of exposures. Spectra are offset vertically for clarity and are measured using a 532 nm excitation laser. The initial Raman spectrum shows peaks at ~101 cm$^{-1}$ and ~138 cm$^{-1}$, corresponding to the $E_g^2$ and $A_{1g}^2$ vibrational modes. (b) The positions and full widths at half max of the $E_g^2$ and $A_{1g}^2$ peaks as a function of argon plasma exposure time. (c) The intensity of the $E_g^2$ and $A_{1g}^2$ peaks, as well as the intensity ratio $I(E_g^2)/I(A_{1g}^2)$ as a function of argon plasma etching time.

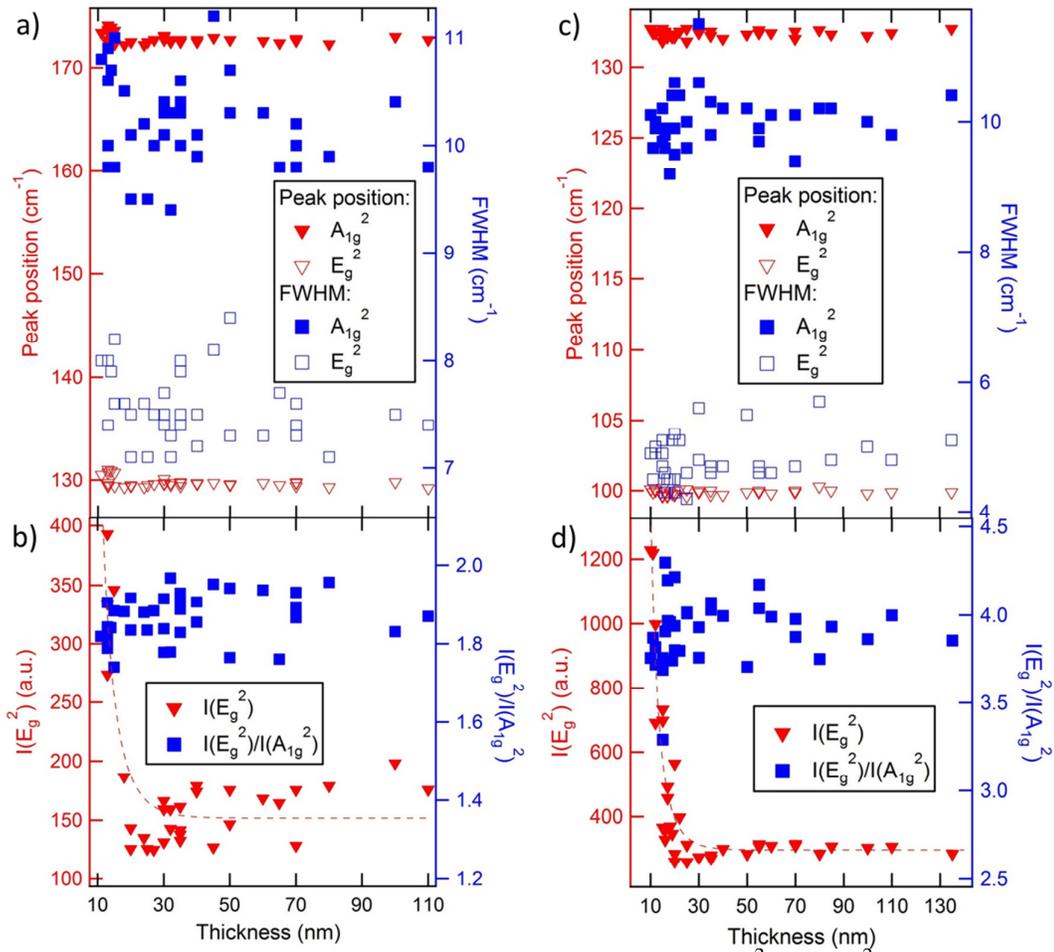

Figure S1. (a) The positions and full widths at half max of the $E_g^2$ and $A_{1g}^2$ peaks as a function of thickness for various non-irradiated exfoliated $Bi_2Se_3$ flakes. (b) The intensity of the $E_g^2$ peak, $I(E_g^2)$, and intensity ratio $I(E_g^2)/I(A_{1g}^2)$ as a function of thickness for the same samples as in (a). (c) The positions and full widths at half max of the $E_g^2$ and $A_{1g}^2$ peaks as a function of thickness for non-irradiated exfoliated $Bi_2Te_3$ flakes. (d) The intensity of the $E_g^2$ peak, $I(E_g^2)$, and intensity ratio $I(E_g^2)/I(A_{1g}^2)$ as a function of thickness for the same samples as in (c). The dashed lines in (a) and (b) represent a multireflection fitting model for thin films.[10]

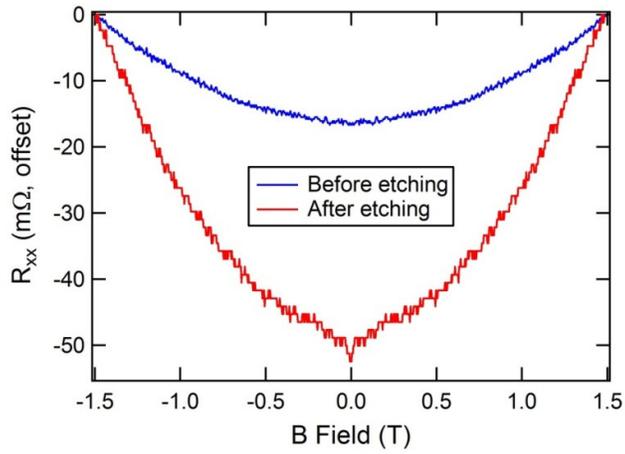

Figure S2. 4-terminal resistance plotted against a variable magnetic field for a ~100 nm thick exfoliated $Bi_2Se_3$ device before and after brief argon plasma etching. The etching did not significantly reduce the thickness of the sample, but still resulted in the emergence of a small weak anti-localization feature around B=0T. The curves are offset for clarity. Before etching, $R_{xx}(B=0) = 4.4\ \Omega$; after etching, $R_{xx}(B=0) = 7.6\ \Omega$.